# Delay-Constrained Utility Maximization in Multihop Random Access Networks

Amir M. Khodaian, Babak H. Khalaj

Department of Electrical Engineering and Advanced Communications Research Institute,
Sharif University of Technology, Tehran, Iran

Email: khodaian@ee.sharif.edu, khalaj@sharif.edu

**Abstract —** Multi-hop random access networks have received much attention due to their distributed nature which facilitates deploying many new applications over the sensor and computer networks. Recently, utility maximization framework is applied in order to optimize performance of such networks however delay is not limited and proposed algorithms result in very large transmission delays. In this paper, we will analyze delay in random access multi-hop networks and solve the delay-constrained utility maximization problem. We define the network utility as a combination of rate utility and energy cost functions and solve the following two problems: *'optimal medium access control with link delay constraint'* and, *'optimal congestion and contention control with end-to-end delay constraint'*. The optimal tradeoff between delay, rate, and energy is achieved for different values of delay constraint and the scaling factors between rate and energy. Different distributed solutions will be proposed for each problem and their performance will be compared in terms of convergence and complexity.

## I. INTRODUCTION

In many wireless ad-hoc networks, due to the lack of a central station, nodes contend for the channel and decide channel access in a random manner. In such networks that channel access is not predetermined and depends on the network traffic, it is possible that two nodes simultaneously decide to send data to the same receiver, resulting in collision. Collisions waste energy, increase transmission delay and reduce throughput. The aim of random access protocols is controlling access to channel (contention control) in order to achieve the desired network performance. One of the main parameters that affect contention is transmission probabilities of the nodes. Another parameter is the arrival rate of traffic at each node that should be

---

This work is supported in part by Iran Telecommunication Research Center (ITRC).

.

controlled properly in order to prevent increasing queue sizes and packet delays. The rate control may be performed either at each node or only at the source of the traffic. These parameters determine the energy consumption, transmission rates and delay of the packets; the three key performance criteria that we will concentrate on optimizing them in this work.

The importance of energy efficiency in ad-hoc networks stems from the multi-hop nature of the network. If a node of an ad-hoc network runs out of energy some routes may become disconnected [1]. Therefore, the available energy of nodes should be consumed cautiously to transmit as much information as possible [2], [3]. Another criterion for the performance of a network is the transmission rate of traffics passing through the network. By combining the transmission rate and energy consumption of different nodes, a *network utility function* can be formed which is a monotonically increasing function of the rate allocated to each source and a decreasing function of the energy consumed at each node. Thus, two separate terms can be considered for the network utility function: the *rate utility function* and the *energy cost function.*

Network Utility Maximization (NUM) has received much attention in the literature [4], [5], [6]. It has been first proposed by Kelly [4] in order to optimize end-to-end rates of the wired networks. It is also used in optimizing transport layer of wireless networks [5] , [6]. Also, Nandagopal et. al. [7] used a similar approach in proportionally fair channel allocation and [8] developed the idea of optimizing persistence probabilities in the random access wireless network. General network utility function for random access and some distributed algorithms with reduced complexity and message passing is also proposed in [9]. Most of the researches in the field only consider the rate or throughput in the network utility function, however, our goal is to also incorporate energy consumption in the network utility function. There are many research activities that their objective is to minimize energy cost function or maximize lifetime for wireless ad-hoc networks [10]-[12] , without considering the rate at the same time.

Transmission delay is another important parameter for the network performance and a delay limit should be practically assumed for the packets in the network. Such delay constraint depends on the type of traffic and the required quality of service (QoS) level. Real-time applications such as voice or video conferencing require packets to be transmitted with a specific delay limit. However, data transmission is less sensitive to delay and a more relaxed delay constraint can be adopted for such transmissions. It should be noted that one of the issues in random access networks is the size of queues, for example, in Aloha the average length of the queues in some nodes may go to infinity [13]. Setting a delay limit for the packet transmission is therefore equivalent to setting a limit for the average queue length. Some recent papers have also addressed the stability of random access networks and proposed algorithms that control transmission probabilities with queue sizes [14], [15]. It is shown that although such algorithms will reduce delay, they cannot guarantee the maximum delay and optimality of them is not verified as well. To the best knowledge of the authors, the work presented in this paper is the first case which considers rate and energy optimization along with delay constraints in multi-hop random access networks. Although [5] and [8] have formulated and solved NUM for the random access model, they have neither considered the energy consumption nor the delay constraint. Also, optimal utility-lifetime tradeoff has been achieved in [12] for non-random access networks, however no delay constraint was considered in that approach. Finally, delay minimization for single-hop slotted-Aloha was considered in [16], however energy minimization and fairness among nodes were not addressed.

In an earlier paper [17], we solved the problem of optimizing network utility (as a function of energy and transmission rate) in random access networks but without any delay constraint. In [18], the notion of delay was added to the context of optimizing random access networks and a delay constraint was set for each link, but end-to-end delay constraint and optimization of the transport layer was not discussed. In this paper, we extend our work in [18] in two ways:

1- Proposing a new distributed solution for the problem of *'optimal MAC with delay constraint'* which is a non-iterative, suboptimal solution shown to be close to optimal.

2- Defining the new problem of *'optimal congestion and contention control with end-to-end delay constraint'* and presenting first and second order distributed solutions for it.

The rest of the paper is organized as follows: The network model and analysis of delay is presented in the next section. Then, in section III we formulate the *'optimal MAC with delay constraint'* problem by defining the goal functions and the link delay constraint and proposing distributed algorithms for solving it. The *'Optimal congestion and contention control with end to end delay constraint'* problem and the required solutions are presented in section IV. Section V, investigates the trade-off between energy, rate, and delay; it also contains numerical results of the distributed algorithms. Finally, we conclude the paper and review its contributions in section VI.

## II. NETWORK MODEL AND ANALYSIS

### A. Network Topology and Definitions

Suppose a slotted-Aloha ad-hoc network consisting of $N$ nodes that transmit their packets through their neighbors using the set of links $L$. Each node selects one of its links and transmits with probability $p_{ij}$ where $i$ is the transmitter index and $j$ is the receiver index. The transmission probability of node $i$, which is equal to the sum of transmission probabilities of its output links is defined by $P_i$. We assume that nodes transmit during time slots whose duration equals the packet transmission time. Nodes are supposed to have infinite buffers such that no packet drop occurs. We also assume that the distribution of packet arrival at each node is Poisson and independent of the other nodes.

The set of neighbors of node $i$ is denoted by $N_i$, and the set of nodes which node $i$ transmits to them and the set of nodes that transmit to node $i$ are denoted by $O_i$ and $I_i$, respectively. In this paper, we assume that all nodes have equal power, resulting in symmetric neighborhoods i.e. $i \in N_j \iff j \in N_i$. The case that neighbors use unequal powers was considered in [17]. Although such assumption can be easily incorporated in the current work, it has not been considered in this paper in order to simplify the formulations.

S denotes set of information sources, where source $s \in S$ uses subset of links, $L(s) \in L$, as a route to transmit its data. We assume no loop exists in the set $L(s)$ and there is a single possible and predefined path from source to destination. The set of sources that share link $(i, j)$ for transmission of traffic is defined by $S(i, j) = \{s \in S | (i, j) \in L(s)\}$ and, therefore, $(i, j) \in L(s)$ if and only if, $s \in S(i, j)$. We use, $packet/slot$, as the unit for transmission rate and also assume that all nodes use the same packet size in the slotted network.

### B. Delay Analysis

In order to calculate average delay in random access networks, we assume packet arrival at each node to be modeled by a Poisson process. It is also assumed that in the case of collisions, the packet is retransmitted until it is successfully received at the other end. Thus, when a packet collides it does not return to the queue, but waits until it is served. We can then model each link as an *M/G/1* queue and use the following Pollaczek-Khinchin formula to estimate the queuing delay [19]:

$$D = W + \bar{S} = \bar{S} + \frac{r\overline{S^2}}{2(1 - \rho)} \tag{1}$$

where $W$ is the waiting time of the queue, $\bar{S}$ is the average service time, $r$ is the arrival rate, and $\rho = r\bar{S}$ is the queue utilization. Calculating delay with (1) requires computation of the first and second order mean of the service time. The service time of each link depends on the transmission probability of that link and the collision probability. In the case of slotted access, the service time is a discrete random variable and the probability of transmission after $k$ time slots is equal to $x(1-x)^k$, where $x$ is the successful transmission probability in each slot. Mean and variance of the service time are then given by:

$$\bar{S} = 1/x \qquad \sigma_S^2 = \frac{1-x}{x^2} \qquad (2)$$

Using (1) and (2) the link delay, in number of slot times, can be found:

$$D = \frac{1}{x} + \frac{r\dfrac{2-x}{x^2}}{2\left(1-\dfrac{r}{x}\right)} = \frac{1}{x} + \frac{r(2-x)}{2x(x-r)} = \frac{(1-r/2)}{(x-r)} \qquad (3)$$

The end-to-end delay of a session can also be computed as the sum of link delays throughout the path of each session.

## III. Optimal MAC with Delay Constraint

Our goal in this section is to optimize MAC parameters in order to achieve minimum energy consumption and maximum rate utility in the network. Solving such a bi-criterion problem is equivalent to finding Pareto optimal points [20]. Pareto optimal points have the characteristics that no other point that is better in both energy and rate utility exists. Also, an additional delay limit for the links of the network is considered in this bi-criterion problem. In problems where the goal and constraint functions are convex, it is common to use scalarization in order to find the Pareto optimal points. However, in this case since the delay constraint in its original form is non-convex, the first step would be to convert it to a convex function. Subsequently, scalarization can be used in order to form a convex problem and achieve Pareto optimal points.

### A. Convex Formulation

The rate utility function, $U_r$, in the MAC layer is defined as the summation of rate utilities over all links. In order to achieve proportional fairness between links we use the same approach as [7] and [8], and define rate utility as a logarithmic function of the link rates:

$$U_r = \sum_{(i,j) \in L} \log\left(r_{ij}\right) \qquad (4)$$

Therefore, the rate utility which is the goal function to be maximized would be concave. Another goal function that should be formulated is the energy consumed in the network. The required energy to transmit a packet by node $i$ is equal to $e_i$, so the average energy consumption of node $i$ transmitting with probability $P_i$ in one timeslot is given by $E_i = e_i \times P_i$. In this paper, we assume equal transmission power for the nodes and thus, total energy consumption of the network is given by the following linear function:

$$E = \sum_{i \in N} E_i = \sum_{i \in N} e_i P_i \qquad (5)$$

The network parameters targeted for the optimization problem are the transmission probabilities and arrival rates. The rate utility and negative of the energy cost are goal functions that should be maximized and were shown to be convex functions of transmission rate and probabilities. The next step is to show that the constraints are also convex functions of these parameters. The link delay constraint, mentioned in section II, can be reformulated as follows:

$$D_{ij} = \frac{(1 - r_{ij}/2)}{(x_{ij} - r_{ij})} < D_c \Rightarrow r_{ij} + \frac{(1 - r_{ij}/2)}{D_c} < x_{ij} \qquad (6)$$

where $D_c$ is the delay constraint for the links and $x_{ij}$ is the successful transmission probability over the link $(i,j)$. Analyzing $x_{ij}$ in non-saturated network is known to be very complex [21]. In order to make the problem tractable we assume that a link will send dummy packets when its queue is empty. Thus, successful transmission probability depends only on transmission probability of the link, receiver and interfering neighbors:

$$x_{ij} = p_{ij}(1 - P_j) \prod_{l \in N_j^{in} - \{i\}} (1 - P_l) \qquad (7)$$

Equation (7) shows that $x_{ij}$ has a product form and is non-convex, so, the delay constraint (6) has non-convex form. In order to obtain a convex delay constraint as a function of $p_{ij}$, we first use a logarithmic function, which is monotonically increasing and preserves the inequality, on both sides of (6):

$$\log(r_{ij} + \frac{(1 - r_{ij}/2)}{D_c}) + \log(1 - P_j) + \sum_{l \in N_j^{in} - \{i\}} \log(1 - P_l) < 0 \qquad (8)$$

It is easy to show that the above constraint is a concave function of the arrival rates $r_{ij}$. Therefore, by using a change of variables of the form $z_{ij} = \log(r_{ij})$, the delay constraint can be converted into a convex function of $z_{ij}$. In this way, the rate utility also changes to a linear function:

$$U_r = \sum_{(i,j) \in L} z_{ij} \tag{9}$$

We can now use the scalarization scheme for the convex goal functions in order to formulate a convex problem and find the Pareto optimal points:

$$
\begin{aligned}
\min \quad & U = \lambda_1 E - \lambda_2 U_r \\
S.t. \quad & \log(\frac{1}{D_c} + e^{z_{ij}}(1 - \frac{1}{2D_c})) - \log(x_{ij}) \leq 0 \\
& 0 \leq P_i, p_{ij} \leq 1 \, \forall i \in N, j \in O_s \\
& P_i = \sum_{j \in O_i} p_{ij} \qquad \forall i \in N
\end{aligned}
\tag{10}
$$

There are many well-known algorithms for solving such convex problems. In section V, we use Sequential Quadratic Programming (SQP) [22] in order to solve (10) and find the optimal tradeoff curves between energy and rate utility for different values of delay constraints.

## B. Feasibility of the Problem

The convex formulation ensures that the problem has a unique solution in its feasible region[20]. Another issue that has to be addressed is feasibility of the problem which depends on the value of the link delay constraint ($D_c$). Therefore, we should find the minimum delay constraint (MinDc), that ensures feasibility of the problem and only adopt higher delay constraints for the network. The delay constraint formula (6) shows that the maximum link delay occurs for the link with the minimum throughput. As a result, if the minimum throughput is maximized over all links, it is possible to obtain the point that can tolerate the MinDc. This is equivalent to the following maxmin optimization problem:

$$
\begin{aligned}
& \max_{\underline{p}} \min_{(i,j)} x_{ij} \\
& 0 \leq P_i, p_{ij} \leq 1 \quad \forall i \in N, j \in O_s
\end{aligned}
\tag{11}
$$

This problem can also be reformulated as the following convex optimization form. The achieved minimum delay constraint would only depend on the network structure.

$$
\begin{aligned}
& \max_{\underline{p}} z \\
& z \leq \log(x_{ij}) \\
& 0 \leq P_i, p_{ij} \leq 1 \quad \forall (i,j) \in L
\end{aligned}
\tag{12}
$$

## C. Distributed MAC Optimization

In general, algorithms such as SQP or Interior Point Methods (IPM) are applicable in a centralized manner. However distributed algorithms are usually the preferred choice in a network as nodes can decide and select their optimal variables through minimum interaction with other nodes. Since the problem is convex and feasible, the duality gap is zero and we can use the dual problem in order to obtain separate problems over the nodes. Such dual decomposition approach will then lead to the corresponding update formulas for link probabilities and rates. First, we write the Lagrangian of (10) as follows:

$$L(\mu, p, z) = \sum_{(i,j)} \left[ \lambda_1 e_i p_{ij} - \lambda_2 z_{ij} + \mu_{ij} \left( \log \left[ (1 - \frac{1}{2D_c}) e^{z_{ij}} + \frac{1}{D_c} \right] - \log(x_{ij}) \right) \right] \tag{13}$$

where $\mu_{ij}$ is the dual variable for delay constraint of link $(i,j)$. Using the derivative of the Lagrangian we can find the rate update formula and the corresponding equation for the link probabilities:

$$r_{ij}^{(n+1)} = \left[ \frac{\lambda_2}{(\mu_{ij}^{(n)} - \lambda_2)(D_c - 0.5)} \right]^+ \tag{14}$$

$$p_{ij}^{(n+1)} \lambda_1 e = \mu_{ij}^{(n)} - \frac{1}{(1 - P_i^{(n+1)})} p_{ij}^{(n+1)} \sum_{\substack{(k,l) \in L \\ l \in N_i \cup i}} \mu_{kl}^{(n)} \tag{15}$$

Using (15) and computing the summation over $j$, results in the quadratic equation (16) for the node transmission probabilities. New link probabilities can then be found using the updated node probability and dual variables:

$$(P_i^{(n+1)})^2 (\lambda_1 e_i) - P_i^{(n+1)} \left( \sum_{\substack{(k,l) \in L \\ l \in N_i \cup i}} \mu_{kl}^{(n)} + \sum_{l \in O_i} \mu_{il}^{(n)} + \lambda_1 e_i \right) + \sum_{l \in O_i} \mu_{il}^{(n)} = 0 \tag{16}$$

$$p_{ij}^{(n+1)} = \text{proj}_{0 < p} \left[ \mu_{ij}^{(n)} \left( \lambda_1 e_i + \frac{1}{1 - P_i^{(n+1)}} \sum_{\substack{(k,l) \in L \\ l \in N_i \cup i}} \mu_{kl} \right)^{-1} \right] \tag{17}$$

where, $proj_{c_1 < x < c_2}$ denotes projection of results over the constraint set $c_1 < \cdot < c_2$:

$$proj_{c_1 < x < c_2} = \max \{c_1, \min \{x, c_2\}$$

By computing the gradient of the Lagrangian in terms of the vector of dual variables, the following formula can be used to update dual variables:

$$\mu_{ij}^{(n+1)} = [\mu_{ij}^{(n)} + \alpha_n \{\log\left[(1 - \frac{1}{2D_c})r_{ij}^{(n)} + \frac{1}{D_c}\right] - \log(x_{ij}^{(n)})\}]^+ \tag{18}$$

Convergence of this dual decomposition algorithm is guaranteed for small values of $\alpha_n$ or when $\alpha_n$ goes to zero for large values of $n$ [23]. It should be mentioned that in our numerical analysis, we have used a constant small step size since such choice does not require synchronous update of the step size in the whole network.

### D. Sub-optimal MAC Problem

We have considered the problem of rate utility optimization without delay constraint in [17] and proposed a non-iterative distributed solution for it. A non-iterative sub-optimal algorithm for the delay constrained problem, (10), can also be proposed by first ignoring the delay constraint and then setting the link rates based on the achieved throughput such that the delay constraint is satisfied:

$$r_{ij} = \frac{D_c x_{ij} - 1}{D_c - \frac{1}{2}}$$

The question that arises here is the performance tightness of the optimal and proposed suboptimal algorithms. For some special cases that delay constraint or throughput is high enough we can suppose $D_c \gg 1/x_{ij} > 1$ and thus:

$$r_{ij} = \frac{D_c x_{ij} - 1}{D_c - \frac{1}{2}} \approx x_{ij} - \frac{1}{D_c} \approx x_{ij}$$

However, since there is no constraint on the minimum throughput of the links in general, this value may become so small that for some values of $(i,j)$, we have $D_c < 1/x_{ij}$ and it becomes impossible to satisfy the delay constraint for such links. In section V, we address the performance of the sub-optimal algorithms through numerical analysis and show that in typical scenarios the sub-optimal algorithm behaves quite similar to the optimal one.

## IV. Cross Layer Optimization with End-to-End Delay Constraint

In this section, we address the end-to-end scenario where the delay of each session is sum of the delay of the links in its path. Based on the delay formulation discussed in section II.B, we first formulate the problem for the end-to-end case and provide a convex formulation for it. The first order and Newton-like algorithms are then provided in order to achieve distributed algorithms.

## A. Problem Formulation

The cross-layer optimization problem considering both the MAC and transport layers is similar to the MAC problem in section III and a multi-objective problem of energy minimization and rate utility maximization is considered. Here, the rate utility function is defined as *sumlog* of session rates which achieves proportional fairness among sources. The tradeoff between energy and rate utility can be controlled using scalars $\lambda_1$ and $\lambda_2$. We assume that each session have a separate delay constraint $D_s$ which depends on its source requirements.

$$\min_{p_{ij}, y_s, r_{ij}} \quad \lambda_1 \sum P_i e_i - \lambda_2 \sum_{s \in S} \log(y_s) \quad \text{(Objective Function)}$$

$$
\begin{aligned}
s.t. \quad & r_{ij} \leq x_{ij} \quad \forall (i,j) \in L \\
& r_{ij} = \sum_{s \in S(i,j)} y_s \\
& y_s \geq 0 \quad\quad\quad \forall s \in S
\end{aligned}
\left.\right\} \text{(Rate and Throughput Constraints)}
$$

$$\sum_{(i,j) \in L_s} \frac{1 - r_{ij}/2}{x_{ij} - r_{ij}} < D_s \quad\quad \text{(End-to-End Delay Constraint)} \tag{19}$$

$$
\begin{aligned}
& \sum_{j \in o_i} p_{ij} = P_i \quad \forall i \in N \\
& 0 \leq p_{ij} \leq 1 \quad\quad \forall (i,j) \in L \\
& P_i \leq 1 \quad\quad\quad \forall i \in N
\end{aligned}
\left.\right\} \text{(Probability Constraints)}
$$

$r_{ij}$ shows the total rate that passes through link *ij*. Since the link throughput $x_{ij}$ that appears in the delay constraint is equal to the service rate of the link and has a product form of (7), problem (19) would not be in the standard convex optimization form. In order to convert it into convex form, first we define $D_{ij}$ as an auxiliary variable for each link and convert the end-to-end delay constraint as:

$$
\begin{aligned}
& \frac{1 - r_{ij}/2}{x_{ij} - r_{ij}} \leq D_{ij} \Rightarrow \log(\frac{1}{D_{ij}} + r_{ij}(1 - \frac{1}{2D_{ij}})) \leq \log(x_{ij}) \; ; \; \forall (i,j) \in L \\
& \sum_{(i,j) \in L_s} D_{ij} \leq D_s
\end{aligned}
\left.\right\} \text{Delay Constraints} \tag{20}
$$

Although the constraint function becomes a linear function of link probabilities ($p_{ij}$), it has the form $\log(ar_{ij}+b)-c<0$ which is not a convex function in its initial form. Therefore by defining $z_s=\log(y_s)$ the following convex optimization problem which has a unique solution in its feasible region is achieved:

$$\min_{z_s,D_{ij},p_{ij}} \lambda_1 \sum P_i e_i^{tot} - \lambda_2 \sum_{s \in S} z_s$$

$$s.t. \quad \log(\frac{1}{D_{ij}} + (\sum_{s \in S(i,j)} e^{z_s})(1-\frac{1}{2D_{ij}})) \leq \log(x_{ij}) \; ; \; \forall (i,j) \in L$$

$$\sum_{(i,j)\in L_s} D_{ij} \leq D_s \qquad \forall s \in S \tag{21}$$

$$\text{Probability and session rate constraints}$$

Note that the first constraint also guarantees the link rate constraint, where arrival rate should be smaller than the service rate:

$$\sum_{s \in S(i,j)} e^{z_s} \leq x_{ij}$$

Problem (21) is in the form of a standard convex optimization problem and can be solved using interior point methods [20] or Sequential Quadratic Programming (SQP) [22]. Such techniques can be used to achieve a centralized solution but in order to obtain a practical distributed solution, an algorithm that gives optimal variables by using the distributed information, should be proposed.

## B. Distributed Solutions

Here, we describe two different distributed algorithms for solving the delay-constrained energy-rate optimization. First, we assume constant transmission probabilities ($p_{ij}$) in (19) and reformulate the problem as follows:

$$\min_{y_s, D_{ij}} \lambda_1 \sum_{i \in N} P_i e_i - \lambda_2 \sum_{s \in S} \log(y_s)$$

$$(\sum_{s \in S(i,j)} y_s)(1 - \frac{1}{2D_{ij}}) + \frac{1}{D_{ij}} \le x_{ij} \quad \forall (i,j) \in L$$

$$\sum_{(i,j) \in L_s} D_{ij} < D_s \quad \forall s \in S \tag{22}$$

$$y_s \ge 0 \qquad \forall s \in S$$

$$D_{ij} > 0 \qquad \forall (i,j) \in L$$

Problem (22) is the delay-constrained congestion control problem which has convex objective and constraint functions. In order to propose a distributed solution for (22), we use a common method of dual decomposition [4][6] and write Lagrangian function for this problem:

$$\mathcal{L}(y, D, \mu, \upsilon) = \sum_{(i,j) \in L} \left[ \lambda_1 p_{ij} e_i + \mu_{ij} \left[ \frac{1}{D_{ij}} + (\sum_{s \in S(i,j)} y_s)(1 - \frac{1}{2D_{ij}}) - x_{ij} \right] \right]$$

$$- \sum_{s \in S} \left[ \lambda_2 \log(y_s) - \upsilon_s \left( \sum_{(i,j) \in L_s} D_{ij} - D_s \right) \right] \tag{23}$$

where, $\mu_{ij}$ and $\upsilon_s$ are Lagrangian variables related to link and session delay constraints, respectively. The Lagrangian can be rewritten and decomposed to achieve the following sub-problems for links and sources:

$$\min_{D_{ij}} D_{ij} \sum_{s \in S(i,j)} \upsilon_s - \mu_{ij}(0.5 \sum_{s \in S(i,j)} y_s - 1) \frac{1}{D_{ij}}$$

$$S.t. \quad D_{ij} > 1 / x_{ij} \tag{24}$$

$$\max_{y_s} \lambda_2 \log(y_s) - y_s \sum_{(i,j) \in L(s)} \mu_{ij}(1 - \frac{1}{2D_{ij}})$$

$$S.t. \quad y_s \ge 0 \tag{25}$$

$\mu_{ij}$ and $\upsilon_s$ can be obtained using an iterative algorithm and in each step (24) and (25) should be solved in order to find session rates and link delays.

$$D_{ij}^{(n+1)} = \operatorname*{proj}_{\substack{1 \\ x_{ij} < D}} \left( \frac{\mu_{ij}^{(n)}(2 - \sum\limits_{s \in S(i,j)} y_s^{(n)})}{2 \sum\limits_{s \in S(i,j)} v_s^{(n)}} \right)^{1/2} \qquad (26)$$

$$y_s^{(n+1)} = \operatorname*{proj}_{0 < y}(\frac{\lambda_2}{\sum\limits_{(i,j) \in L} \mu_{ij}^{(n)}(1 - \frac{1}{2D_{ij}^{(n)}})}) \qquad (27)$$

Now, we propose two different methods to find Lagrangian variables and transmission probabilities. First, we use projected gradient scheme to achieve the following set of update formulas:

$$h_{ij}^{(n)} = (1 - \frac{1}{2D_{ij}^{(n)}})(\sum_{s \in S(i,j)} y_s^{(n)}) + \frac{1}{D_{ij}^{(n)}} - x_{ij}^{(n)}$$

$$\mu_{ij}^{(n+1)} = \left[ \mu_{ij}^{(n)} + \alpha_n h_{ij}^{(n)} \right]^+ \qquad (28)$$

$$g_s^{(n)} = \sum_{(i,j) \in L(s)} D_{ij}^{(n)} - D_s$$

$$v_s^{(n+1)} = \left[ v_s^{(n)} + \beta_n g_s^{(n)} \right]^+ \qquad (29)$$

where, $[x]^+$ denotes $\max(0, x)$ and, $\alpha_n$ and $\beta_n$ are update coefficients which should be small positive variables in order to ensure proper convergence.

Transmission probabilities can also be updated in the gradient direction. The sensitivity theorem [23] states that the gradient of the objective function at the optimal point relative to a variable constraint is equal to the Lagrangian vector of that constraint. Thus, the gradient of the objective function of (22) with respect to link throughputs would be equal to the vector $(\dots, \mu_{ij}, \dots)$. By solving (22), the transmission probabilities can be updated in the gradient direction:

$$f_{ij}^{(n)} = \lambda_1 e_i - \sum_{(s,t) \in L} \mu_{st} \frac{\partial x_{st}(\mathbf{p}^{(n)})}{\partial p_{ij}}$$

$$p_{ij}^{(n+1)} = \operatorname*{proj}_{0 < p < 1}(p_{ij}^{(n)} - \varphi_n f_{ij}^{(n)}) \qquad (30)$$

Also, as node transmission probabilities $P_i$ should be projected on $P_i < 1$, all link probabilities should be reduced when this constraint is violated but they should also remain non-negative.

Although we may increase convergence rate using the Newton algorithm, such approach requires computing the Hessian matrix which cannot be calculated with local information [24]. Thus, we propose a Newton-like algorithm which employs the diagonal elements of the Hessian and has supper-linear convergence rate. Applying this algorithm to solve problem (22), the following update formulas will be achieved for Lagrangian variables:

$$H_{ij}^{(n)} = \frac{h_{ij}^{(n)} - h_{ij}^{(n-1)}}{\mu_{ij}^{(n)} - \mu_{ij}^{(n-1)}}$$

$$\mu_{ij}^{(n+1)} = \left[ \mu_{ij}^{(n)} + \alpha_n h_{ij}^{(n)} \big/ H_{ij}^{(n)} \right]^+ \tag{31}$$

$$G_s^{(n)} = \frac{g_s^{(n)} - g_s^{(n-1)}}{v_s^{(n)} - v_s^{(n-1)}}$$

$$v_s^{(n+1)} = \left[ v_s^{(n)} + \beta_n g_s^{(n)} \big/ G_s^{(n)} \right]^+ \tag{32}$$

Also, we use the same Newton-like method to update transmission probabilities and obtain the following iterative formulation:

$$F_{ij}^{(n)} = \frac{f_{ij}^{(n)} - f_{ij}^{(n-1)}}{p_{ij}^{(n)} - p_{ij}^{(n-1)}}$$

$$p_{ij}^{(n+1)} = \text{proj}_i \left( p_{ij}^{(n)} - \varphi_n f_{ij}^{(n)} \big/ F_{ij}^{(n)} \right) \tag{33}$$

In section V, we will compare the convergence rate of the gradient projection and the Newton-like algorithms through numerical results.

## V. Numerical Analysis

In our numerical analysis, we have mainly used the sample network of Fig. 1, containing 10 nodes, 12 links, and 4 sessions. We should note that numerical results are obtained using MATLAB and its optimization toolbox.

*A. Centralized Solution for MAC Optimization*

As mentioned in section III, the minimum delay constraint (*MinD$_c$*) of the links is a parameter that should be properly computed by solving (12) to guarantee that the selected delay constraint will not result in an infeasible problem. For the sample network of Fig. 1.(a), solving (12) has led to a *MinDc* value of **10.47**. Also, in order to understand effect of network topology on *MinDc* we have considered two special cases of linear and star networks with $n$ nodes and $(2n − 2)$ links (Fig. 1.(b) and Fig. 1.(c)). Linear and star networks can be scaled in order to investigate the problem for different contention environments and network sizes. As shown in Fig. 2, for linear networks this minimum delay changes very slowly with the network size; however, for the star network it linearly increases as the number of nodes increases. This clearly provides a rule of thumb for analysis of complex networks and to estimate *MinDc* based on the maximum degree of the nodes in the network.

The cost function of problem (10) is a linear combination of energy and rate utility. The parameters $\lambda_1$ and $\lambda_2$ can be changed in order to control the tradeoff between energy minimization and rate utility maximization. We also use the delay constraint of about $4\times$ *MinDc* or higher in order to obtain a large enough feasible region. This allows $\lambda_1$ and $\lambda_2$ to conveniently control the tradeoff between energy and rate utility over the feasible region. Fig. 3 shows the tradeoff between energy and rate utility for three different delay constraints in case of the sample network. As the delay constraint becomes more relaxed, the optimal network consumes less energy and achieves higher rate utility. For example, our numerical analysis shows that increasing delay from 40 to 100 is more effective than increasing it from 100 to 1000. Also, three regions from left to right can be distinguished on each curve. At large values of $\lambda_1$, the energy is close to its minimum value and changes very slowly, but the rate utility decreases at a high rate. In the next region, the tradeoff between energy and rate utility is more evident. The last region is where rate utility slowly reaches its maximum value at the cost of doubling the energy consumption.

*B. Distributed Algorithm for MAC Optimization*

A simple distributed algorithm that was described in section III.C is used for the case of $(\lambda_1, \lambda_2) = (5, 0.1)$ and a delay constraint equal to 100. The algorithm starts from initial transmission probability of 0.1 for all links and results in the convergence characteristics shown in Fig. 4 where the percentage of error in network cost function, transmission probability, and rate of the link ($i = 1, j = 2$) are plotted for every iteration. Note that for iterative algorithms, the error percentage of variable $x$ in each iteration is defined as $|x(itr) - x^{opt}|/|x^{opt}|$. If we use an error of less than 1% as a measure of convergence, it can be verified that the distributed algorithm converges in about 12 iterations for the sample network.

One interesting question to address is how the convergence of this algorithm scales with the network size. In order to investigate this effect, a linear network has been considered (Fig. 1.(b)). The convergence rate is then compared for different network sizes in which $D_c$=100 and $(\lambda_1, \lambda_2) = (5, 0.1)$. Our numerical analysis shows that for all linear network sizes between 4 and 32, the number of iterations required for convergence is roughly constant and is about 15. In order to explain this, we note that computations of the gradient direction methods scale with the dimension of the problem. However, when distributed computation is used, the number of nodes (computers) also scales with the network size. Consequently, the number of iterations computed by all nodes in finding the optimum point does not scale with the network size.

We also proposed a suboptimal distributed algorithm in section III.C. This suboptimal algorithm is applied to the sample network and comparison with the optimal solution of MAC optimization problem in Fig. 5 shows that performance of the suboptimal algorithm is tight to the optimal one.

## C. Centralized Solution for Cross-Layer Optimization

The Cross-layer optimization problem (19) was converted to a convex form in (21) and solved for the sample network for different values of $D_s$ (session delay constraints), $\lambda_1$, and $\lambda_2$ in order to attain the optimal tradeoff between energy, rate and delay as illustrated in Fig. 6. It is assumed that the same delay constraint is applied for all of the sources. It shows that relaxing delay constraint higher than 800 slot times will not effectively change the rate utility and energy consumption. Similar to the MAC problem three regions from left to right can be distinguished on each curve where energy changes slowly at the first region and rate utility changes slowly for the last region.

## D. Distributed Algorithms for Cross-Layer Optimization

The performance and complexity of gradient projection and Newton-like distributed solutions for the cross-layer problem (19) will be compared in Fig. 7. We have assumed $(\lambda_1, \lambda_2)$=(0.005,10) and an equal delay constraint of 100 packet time for all sessions. The error percentage of the iterative algorithm is computed for the rate of session 1 $y_1$, transmission probability of link ($i$=5, $j$=6), and the rate utility function, as plotted in Fig. 7 (a) and (b). As can be verified, the convergence rate of the Newton-like algorithm is about 10 times faster than the gradient projection, however as equations (28)-(33) show, the Newton-like algorithms require one division per update and also an additional memory for storage of the primal and dual variables of the previous iteration.

## VI. CONCLUSION

The main contribution of this paper is to add the delay constraint to the previous work on network utility optimization in random access networks. We have modeled links as M/G/1 queues and used this model in order to calculate the average delay of the random access protocol. Based on the analysis and by defining the network utility as a function of rate and energy, two related problems are formulated: 'optimal MAC with link delay constraint' and 'optimal contention and congestion control with end-to-end delay constraint'. Both of the problems are reformulated as standard convex problems. Solution of the convex problem is not only useful for network design but can also be used to find optimum achievable energy and rate values at different delay constraints. Our numerical analysis shows that a tradeoff between energy, rate and delay in the random access network both for the MAC and cross-layer problems can be achieved.

For the MAC problem, the minimum delay constraint that ensures feasibility of the optimization problem, *MinDc*, is defined and it is shown that a convex maxmin problem should be solved in order to find *MinDc*. Numerical results indicate that this value scales linearly with the maximum degree of the network nodes. The optimal and suboptimal distributed algorithms are also presented for the MAC problem. The main advantage of the non-iterative suboptimal algorithm is that it does not require a large amount of messages passing between nodes. Although we have shown that the suboptimal algorithm may not work in some special cases, it performs close to optimal in most of the circumstances.

The second part of the paper focuses on the cross-layer problem of MAC-Transport optimization where the end-to-end delay constraint which is non-convex and non-separable in the original form is also taken into account. The problem is turned into a convex problem in order to provide a centralized solution and subsequently the sensitivity theorem is applied to achieve distributed solutions. The Newton-like distributed solution for this problem is thus proposed and shown to be faster than the gradient projection algorithm and thus reduces number of messages passing through the network for optimization of the network.

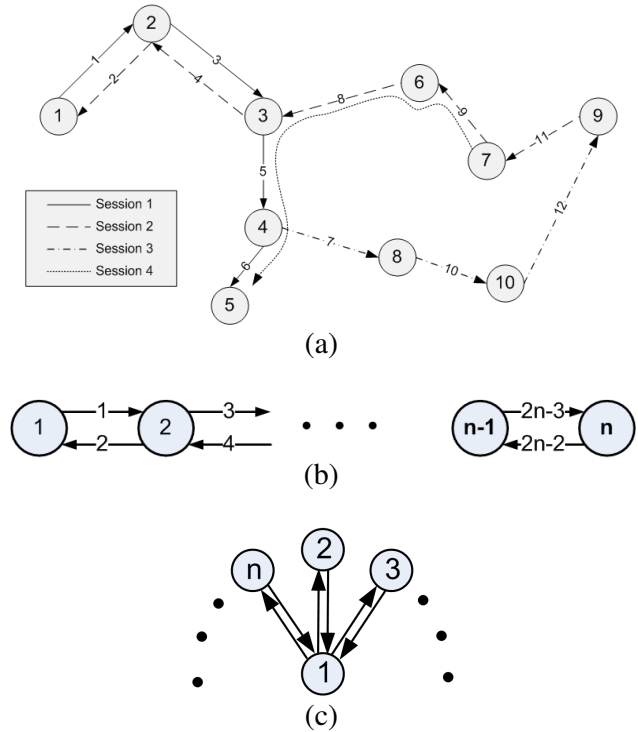

(a)

(b)

(c)

**Fig. 1    Network topologies (a) Sample network (b) Linear network
(c) Star network**

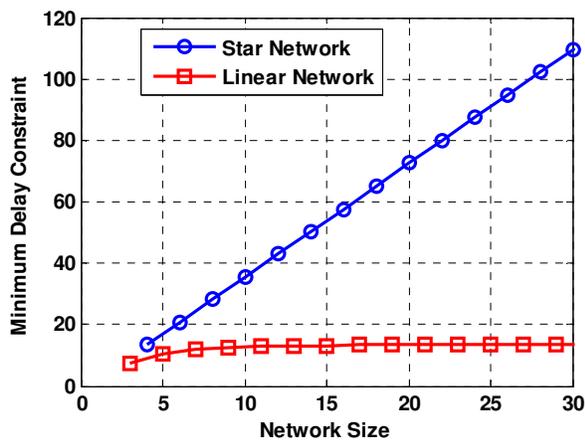

**Fig. 2    The value of the minimum delay constraint
for star and linear networks**

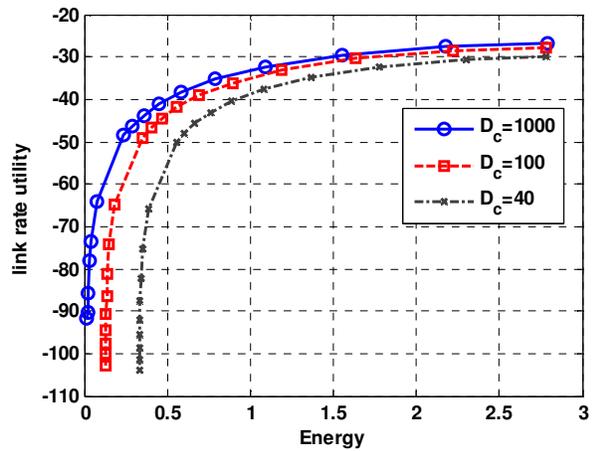

**Fig. 3    The optimal energy-rate utility tradeoff for
different value of delay constraints**

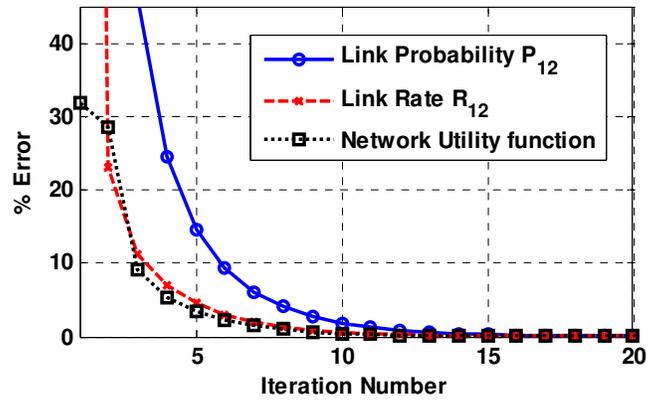

**Fig. 4 Convergence of the distributed MAC optimization algorithm.**

**The percentage of error in $x$ is defined as $|x(itr)-x^{opt}|/|x^{opt}|$**

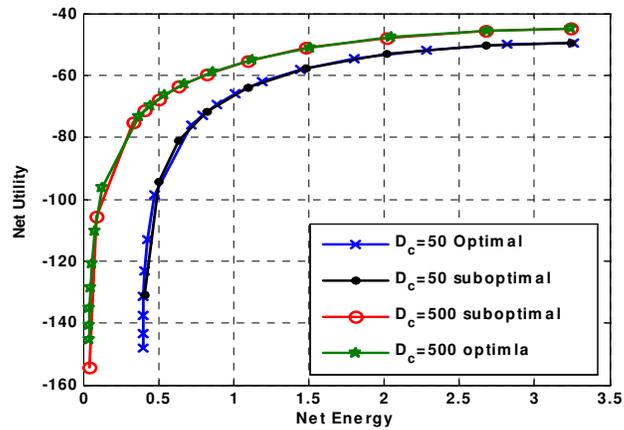

**Fig. 5 Comparison of the sub-optimal and optimal MAC optimization**

**algorithms**

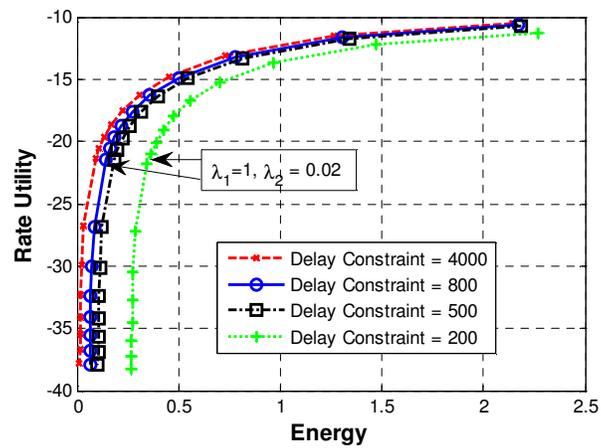

**Fig. 6 Energy-rate utility-delay tradeoff for sample network**

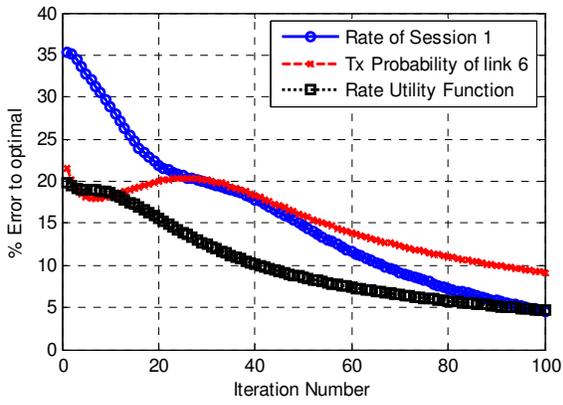
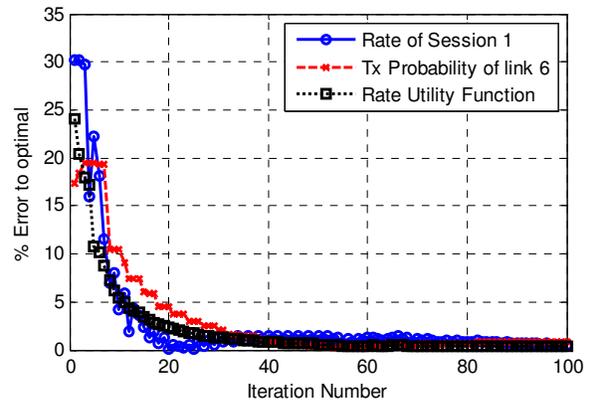

**Fig. 7    Convergence of distributed cross-layeroptimization algorithms**

**(a) gradient projection (b) Newton-like**